\documentclass[12pt,a4paper]{article}
\usepackage[utf8]{inputenc}
\usepackage{amsmath}
\usepackage{amsfonts}
\usepackage{amssymb}
\usepackage[english]{babel}

\usepackage{color}
\usepackage{epsfig}
\usepackage{cite}
\usepackage{graphicx}%
\usepackage{dcolumn}%
\usepackage{bm}%

\begin{document}
	
	\begin{center}
{\large Stability of Gluonic Systems with Multiple Soft Interactions} \\
\vspace*{1cm}
\author{}Rahul Kumar Thakur\textsuperscript{1}, Bhupendra Nath Tiwari\textsuperscript{2,3} and Rahul Nigam\textsuperscript{1}\\
\vspace*{1cm}
$^1$ BITS Pilani, Hyderabad Campus, Hyderabad, India.\\
$^2$ INFN-Laboratori Nazionali di Frascati,	Via E. Fermi 40, 00044 Frascati, Rome, Italy.\\
$^3$ University of Information Science and Technology, ``St. Paul the Apostle", Partizanska Str. bb 6000 Ohrid, Republic of Macedonia.\\
\vspace*{1cm}
	\end{center}
	\begin {abstract}
In this paper, we investigate the stability properties of soft gluons in SIBYLL 2.1 with reference to its original version 1.7 that corresponds to 
hadronic hard interactions. In order to investigate the stability structures, we classify the regions of the gluon density fluctuations in its double leading logarithmic approximation and its equivalent description as the fractional power law. In the parameter space of initial transverse momentum $Q$ and QCD renormalization scale $L$ that correspond to extensive air showers of cosmic rays, we have categorized the surface of parameters over which the proton is stable. We further discuss the nature of local and global correlations and stability properties where the concerning statistical basis yields a stable system or undergoes a geometric phase transition.
Finally we give a phenomenological understanding towards the stability of soft interactions, Pomeron particle productions in minijet model, string fragmentation and verify our result corresponding to the  experiments - CDF, P238, UAS, GEUS and UA4 collaborations.
\end{abstract}

\section{Introduction}

Cosmic rays interactions when viewed as high energy fixed target collisions, involve heavy particles with low intensity. Such interactions are explored indirectly through Extensive Air Showers (EAS) Hiroaki {\em et al.}(2017). In the atmosphere, cosmic ray interactions are considered as the high energy fixed collisions involving various heavy particles. Because of their low intensity, cosmic rays with energies above $10^{15}$ eV can only be studied indirectly through the EAS as they are produced in the atmosphere. The analysis of EAS relies on air shower Monte Carlo simulations which are used hadronic interaction models. At higher energies, where the cosmic ray energy is beyond the reach of man-made accelerators, hadronic interaction properties have to be extrapolated. The difficulties in the extrapolation are augmented by the fact that, while the forward region contains most of the accelerator measurements are made in the central region. On the other, the exploration of the cosmic ray energy spectrum and its compositions for energies greater than about $10^{14}eV$  are of prime interest, as well. An indirect analysis of the data from EAS is interpreted in terms of the above mentioned spectrum and composition solely through the use of accelerator data.

Analysis of such air showers is performed via Monte Carlo simulations using an effective model of hadronic interactions, see Wang {\em et al.}(1991) towards multiple jet production in the light of pp, pA, and AA collisions. In high energy phenomena, the known properties of hadronic interactions which have emerged via existing accelerators have to be extrapolated to ultra high energies ($\ge$ 1 TeV). For such cases, most of their energetics is contained in the forward region as far as shower developments are concerned. Accelerator measurements of the interactions happening in the central region include multiple soft interactions, parton density fluctuations, diffraction dissociation etc. These measurements exhibit a better agreement with fixed target collider data Alekhin {\em et al.}(2018).

With the above motivations, we carry out an intrinsic analysis towards the understanding of the interactions happening in the forward region in the limit of SIBYLL 2.1. Specifically, we  focus on fixed target hadronic interactions, collider experiments, dual parton model (DPM) Capella {\em et al.}(1994) and subgluonic interactions.  The cosmic ray interaction event generator SIBYLL is widely used in extensive air shower simulations. Hereby, we study the underlying statistical fluctuation properties through the gluon density fluctuations in the light of extensive air showers of cosmic rays. Our analysis follows from the minijet model involving hard interactions using a steeply arising gluon density as a function of the model parameters, viz. the initial transverse momentum $Q$ of the colliding particle and QCD renormalization scale $L$ Muta {\em et al.}(2009). To understand the hadronic interactions, we consider parametric distribution where the hadrons-nucleus interactions are analyzed via Glauber scattering theory Glauber {\em et al.} (1970) and nucleus-nucleus interactions via semisuperposition model Engel {\em et al.}(1992). Hereby, the string fragmentation involving quark-antiquark and quark-diquark pairs in the DPM picture arise via air shower simulations in the light of cosmic rays Alekhin {\em et al.}(2018).

In this paper, we concentrate on the stability analysis of such interactions for a varied range of energy scales. Note that the neutral pion equation arises above $10^{19}$ eV, which explains the small lifetime of these pions. It is worth mentioning that our model based on SIBYLL 2.1 is well suited for understanding proton-proton scattering, parton structure functions, multiplicity fluctuations of charged particles in high energy diffractive events and soft interactions as mediated by an event generator Engel {\em et al.} (1999). With this perspective, our description of soft interactions and diffraction dissociation is intrinsically contained in parametric fluctuations of the gluon density. Hereby, we have performed an analysis of gluon density fluctuation in order to understand hadron-hadron/nucleus interactions at high energies, particularly in the forward region. Namely, we primarily look for the parameter space that lead to stable secondary proton which is obtained from the study of the stability of gluon density under variations of its model parameters. 

Following the minijet setup of the phase space containing QCD improved parton whose transverse momentum satisfies the energy dependent relations, we explore the stability of quark-quark and quark-diquark pairs connected via an effective string. Here the sampling of the momentum fraction is done via fractional energy distribution of the quark. In this concern, the fractional energy distribution of diquark is obtained via the corresponding fractional energy distribution of quark within effective quark mass. To consider stability property of multiple soft interactions, one uses Regge theory with a dependence of the impact parameter in contrast to the hard interactions, where the Good-Walker model (Good {\em et al.}1960,Fletcher {\em et al.}1994)plays an important role as a two channel eikonal model. 

Interestingly, our model is well applicable for low and high mass diffraction dissociation despite the fact that they are not well understood in the conventional gauge theory of hadronic interactions and collider physics. Multiple soft interactions as well as new partons yield large multiplicity at high energy, whereby diffraction dissociation phenomena offer an improved understanding of multiplicity distribution of such particles. Herewith, we offer an intrinsic analysis of hadronic interactions towards the understanding of their forward region where most of the particles are expected to be found in the light of the extensive air shower developments.

In the above description of the gluon density function as a function of the fractional energy and transverse momentum mediated by small coupling constant satisfying a geometric saturation relation, we explore an ensemble of QCD configurations in order to determine the statistical stability of a proton in the setup of a minijet model, which holds beyond the conventional collinear factorization approximation Lipatov {\em et al.}(1997). As far as phenomenological events are concerned, the output of our analysis matches well the with collisions mediated via soft interactions and this helps in the understanding of deviations of theoretical predictions with their corresponding data counterparts. In particular, the behavior shown by transverse momentum as a function of energy varies in accordance with the predictions of SIBYLL 2.1.

The effect of gluon density fluctuations on physical observables such as the multiplicity and pseudo-rapidity distribution is described as follows: 
Both the fixed target experiments and collider counterparts provide valuable information in the modeling of hadronic interactions. Namely, the fixed target experiments yield data in the forward region of interactions that are most relevant towards the study of cosmic rays. However, the energies that are relatively lower than the laboratory frame energy $E_{lab}$ are of the order of several hundred GeV. Namely, in collider experiments, one can probe such high energies with the laboratory frame energy $E_{lab} \sim 10^6 GeV$. Following the same, most of the information is collected for the central region of collisions. Nonetheless, there are experiments such as H1 and ZEUS Breitweg {\em et al.} (2002) that are capable of detecting the forward region events. Namely, the experiments such as LHCf and TOTEM  Ahn {\em et al.}(2009) in the LHC at CERN are supposed to collect data in the forward region at energy scales that are equivalent to cosmic rays, viz. $E_{lab} \sim 108 GeV$. 

These experiments produce more particles with an augmented distribution in momentum space that are evident when comparing the associated central region data. In nutshell, both the versions SIBYLL 2.1  and SIBYLL 1.7 have offered a good fit to the rapidity and Feynman distribution for charge particles for pp-collisions in fixed target experiments such as NA49 Gornaya {\em et al.}(2018). It is worth mentioning that the changes made in SIBYLL 2.1  are very evident in the central region, namely, one observes a distinct change in the pseudorapidity and the overall multiplicity distribution of charged particles Ahn {\em et al.}(2009). As a result, in SIBYLL 2.1, the air showers are quickly developed with smaller shower maximum and larger muon number than its former counterpart. A comparative analysis of SIBYLL 1.7 and SIBYLL 2.1 is given in Section 2.1 below.

In the light of cosmic rays air showers, the transverse momentum and renormalization scale on the multiplicity and pseudo-rapidity distribution play an important role in selecting the forward region of interactions. In this concern, we have selected a forward region of interactions by properly choosing the range of the Mandelstam variables as the energy of the constituent quark and diquark pairs. At higher energies, where the energy of cosmic rays is beyond the reach of a man-made accelerator, the hadronic interactions are understood by extrapolation techniques Ahn {\em et al.}(2009). In such extrapolations, the forward region contains most of energetic particles that support the formation of shower developments of cosmic rays. However, due to the string fragmentation involving quarks and diquarks, the difficulties are increased with the fact that the most of accelerator measurements are performed in the central region Ahn {\em et al.}(2009), whereby one obtains a saturated gluon density profile in its the double leading-logarithmic approximation. In this paper, we study fluctuations of the corresponding gluon density in the space of model parameters that leads to stable secondary protons.

In the light of the strong interactions, there exist recent  investigations towards the QCD phase structures, their decoding and particle production at high energy Andronic {\em et al.} (2018).  
This study is based on the lattice Monte Carlo simulations of the phases of QCD. At high temperature, it is anticipated that there is a phase change from a confined hadronic matter to a deconfined quark-gluon-plasma, where the quarks and gluons can travel distances that critically exceed the size of hadrons (Andronic {\em et al.}2018, Andronic {\em et al.}2018). Following the same, it is widely believed that quantum chromdynamics should undergo a phase transition at a finite temperature and/ or baryon density (Kalashnikov {\em et al.}1979, Bellwied {\em et al.}1994). 
Furthermore, the chiral symmetry should be restored and the quarks should be deconfined at a sufficiently high temperature with a definite baryon chemical potential (Stephanov{\em et al.}2006,Meyer{\em et al}1996,Cheng{\em et al.}2006). 
Perspectives phase transitions (or more generally, the QCD thermodynamics) are essential tools for an understanding of the underlying physics of early universe, as well as, the core of neutron stars and associated laboratory experiments at relativistic heavy ion collisions (Rapp {\em et al.}2002,Bass{\em et al.}1999, He {\em et al.}2007).
With this motivation, we explore the stability structures of gluonic systems with multiple soft interactions in the light of background color fluctuations in minijet models. Namely, we analyze gluon density fluctuations in  double leading logarithmic approximation to understand extensive air showers of cosmic rays.
Prospective study of its next version and a comparative analysis of LHC data are left open for future research.

The aim of this paper is to offer a statistical understanding of the stability of hadronic interactions and their associated phase transition phenomena satisfying double leading logarithmic approximation in terms of the gluon density function with an identical Bjorken scaling and transverse momentum of the scattered partons. In this concern, one has a steeply increasing gluon density at fractional energy as exhibited in the results of HERA (Adloff {\em et al.}1997, Breitweg {\em et al.}1997). Note that parton densities in the limit of SIBYLL 2.1 discussed in GRV data (Gluck {\em et al.}1995, Gluck {\em et al.}1998) shows that gluon density as a function of fractional energy $x$ scales as $x^{-(1+\delta)}$ where $0.3 \le \delta \le 0.4$. In this regard, we provide a brief overview of the gluon density function in section 2 and associated fluctuation theory perspective section 3. In the sequel, in section 4, we provide fluctuation theory analysis of the gluon density fluctuations with their qualitative discussion in section $5$. In section 6, we give the conclusion of our analysis with perspective direction for the study of random observation samples.

\section{Gluon density function: an overview}

In this paper, we study statistical fluctuations of gluon density in the limit of double leading logarithmic approximation that is well suited for study of hadronic collision and in particular the soft interactions. 

\subsection{SIBYLL event generators}

First of all, the most important changes in the version SIBYLL 2.1 are that it provides the description of soft interactions and diffraction dissociation, while its former version SIBYLL 1.7 deals only with hard interactions. Namely, in order to allow multiple soft interactions as in SIBYLL 2.1 , the eikonal for the soft interactions is described by using Regge theory, whereas in SIBYLL 1.7 version the eikonal for the hadronic interactions was energy independent and it had the same impact parameter dependence as used for hard interactions. Hereby, the distribution of the secondary particle multiplicity, energy variation of multiplicity, the distribution of secondary particles at high x values and others depending in the soft interactions will change. Namely, in version SIBYLL 1.7, the cross section for diffraction dissociation is parameterized independently of the eikonal of the model, that is, a two-channel eikonal model based on  Good-Walker model, see Good {\em et al.}(1960) for the associated fundamentals of SIBYLL 1.7 and SIBYLL 2.1.

Gluons are the glue-like particles that bind the quarks within the baryons and mesons Mutta {\em et al.}(2009).In Quantum Chromodynamics (QCD) Muta {\em et al.}(2009) high energies fluctuation effects arise when the Pomeron loop equations are used to describe the dipole evolution with an increasing rapidity Hiroaki {\em et al.}(2017) as an effective field theory. In the kinematical range, the description of such an QCD-improved  dynamics of colliding particle which will be probed in the future electron hadron colliders is still an open question. However, the corresponding phenomenological investigations indicate that the gluon number fluctuations are related to discreteness in the QCD evolution that are negligible at the energy scale of HERA Sarkar {\em et al.}(2018). Here, the magnitude of such effects play an important role in the next generation colliders Amaral {\em et al.}(2014). In the realm of QCD, it is known that the gluon gluon interactions dominate the production of bottom quarks at hadron collider energies. On other hand, the gluon-quark interactions interpolate inclusively to prompt photon production at a large transverse momentum in pp collisions at fixed-target energies Berger {\em et al.}(1992)

The underline uncertainties of the gluon density as extracted from the global parton model is large enough in the kinematical range when the Bjorken variable takes a small value at the scale of hard interactions Gonçalves {\em et al.} (2016).The early measurements have revealed a compelling evidence for the existence of a new form of nuclear matter at extremely high density and temperature. This yields a medium in which the predictions of QCD can be tested, whereby new phenomena can be explored under conditions where the relevant degrees of freedom over a given nuclear volumes are expected to arise from the interactions of quarks and gluons rather than that of hadrons.  This lies in the realm of the quark gluon plasma, where the predicted state of matter whose existence and properties are under exploration at the RHIC experiments Arsene {\em et al.} (1978)

\subsection{Gluonic matter formation} 

In the setup of SIBYLL 2.1, the parton distribution functions (PDF) of the proton are essential in order to make theoretical predictions and potentially obtaining developments towards the physical understanding of the experimental results arising at high energy colliders. An accurate determination of such PDFs and the undermining uncertainties lying in the global analysis are therefore crucial. There have been various considerations in this direction by several groups of researchers,(Alekhin{\em et al.}2010,Sjostrand{\em et al.}2006,Pumplin{\em et al.}2009,Aaron{\em et al.}2010,Khoze{\em et al.}2018) In this paper, we offer theoretical predictions towards the stability of gluon density function under the variation of its model parameters and compare the results with the existing experiments. In addition, we offer the prediction towards the stability of gluonic matter formation.

As far as the deep-inelastic scatterings (DIS) are concerned, the hard proton-proton collisions or proton-antiproton collisions at high-energies happen via the scattering of partonic constituents of hadrons. To predict the rate of these processes, a set of universal parton distribution functions would be required. Such distributions are well determined by a global fit to the existing results via DIS and associated hard-scattering data Martin {\em et al.}(2009). At ultrahigh energies, the cosmic ray interactions in the atmosphere can be explored with fixed target collisions involving heavy particles. Because of their low intensity, cosmic rays with energies above $10^15$ eV can solely be studied through the extensive air showers (EAS) which analyses such interactions in the atmosphere Ahn {\em et al.}(2009).

Herewith, one of the central predictions of QCD is the transition from the confined phase that is chirally broken to the deconfined phase which is chirally symmetric state of quasi-free quarks and gluons. The above setup is termed as quark-gluon plasma (QGP) which has been explored via heavy ion colliders (Heinz{\em et al.}(2000).Goncalves{\em et al.}2002) While the high energy hadronic interactions studied at the LHC involve QCD effects that are yet to be well understood as they remain largely unexplored experimentally. It would be interesting to push further the above study of soft interactions at ultrahigh energies in order to detect the cosmic rays in the atmosphere. 

\subsection{Ultra high energy colliders}
At this juncture, by extrapolating Monte Carlo simulations to ulta high energies, one often resorts to understand such phenomenological models (Sjostrand{\em et al.}2006, Bahr {\em et al.}2007, Ranft {\em et al.}1995, Roesler {\em et al.}2001, Engel {\em et al.}2001, Drescher {\em et al.}2001, Ostapchenko{\em et al.}2001, Kalmykov{\em et al.}1979, Sciutto{\em et al.}2001,Engel{\em et al.}1992, Fletcher{\em et al.}1992). These studies are largely based on fits to associated lower energy data. For a given QCD renormalization scale, our interest lies on contributions emerging from jets with a given initial value of the relative transverse momentum in the light of minijet or semihard jet models. To understand the internal structure of nucleons, the nucleon-electromagnetic form factors are used in the existing models Pankaj{\em et al.}(2015) Fundamentally, the internal structure of the nucleon involve the electromagnetic Dirac and Pauli form factors that are equivalent to the electric and magnetic dipoles. Such form factors are examined at the level of experiments and phenomenology  Arringtin{\em et al.}(2007).

In particular, in order to test the stability of gluonic matter formation, calculus based tools are used to find the maxima, minima and saddle point of the gluon density function. Namely, we carry out the analysis of the density profile function in the double leading logarithmic approximation in order to understand the hadron-hadron interactions. The fluctuations are examined in terms of the model parameters. Thereby, we are interested in finding the region of the parameter space where a proton is stable. In the next section, we discuss such regions in detail with the associated analysis of the formation of metastable particles like Pomerons, Reggions and other particles that may be detectable with a small probability. 

It is worth mentioning that the SIBYLL 2.1 retains the DPM picture of hadronic interactions. Namely, in the DPM picture, a nucleon is composed of quarks in particular as a color triplet, and diquarks as a qq-color antitriplet. Herewith, the soft gluons are exchanged via an interaction and the color fields get reorganized as the interactions proceed. In such a model, the projectile quark or diquark combines with the targeted diquark or quark. Such a pair of quarks and diquarks is connected by a pair of strings. In due course, each string fragments separately via the Lund’s string fragmentation model Engel {\em et al.} (1999) Scattering processes at high energy corresponding to the hadron colliders are classified as either the hard or soft interactions. Hereby, it follows that the QCD plays a vital role in understanding the underlying theory of such processes. The methodology and level of the understanding vary in both the cases. For hard processes such as the Higgs boson or high transverse momentum jet production, the rates and associated stability properties are studied by using perturbation theory.

\section{Review of the proposed model}

In this section, we offer an overview of soft interaction in order to discuss their stability properties. We are interested in studying hadron-hadron collisions by invoking the role of gluon density and associated QCD improved parton models  Arringtin {\em et al.}(2007) form factors Heikki {\em et al.}(2016) and jet quinching Amaral {\em et al.}(2014) We study optimization of hard/ soft interactions by using a steeply rising gluon density function corresponding to impact parameter dependent soft profile function of a proton. Given the soft profile function for a proton-proton collision as in Hiroaki {\em et al.}(2017) the model parameters are the QCD renormalization scale $L$ and initial transverse momentum of the colliding particle $Q$. Here, the value of parallel component plays the role of the soft-hard function and the transverse component determining the effective radius of the proton. 

Notice that the choice of the parameter $L$ and $Q$ yields a hard profile in the limit of SIBYLL 1.7. Here, the parameters $\{L, Q \}$ represent a two dimensional surface in the ensemble space of QCD-improved models of partons.  On the other hand, the choice of the parameter $L,Q$ depends solely on the geometric saturation condition of the colliding  profile function of the quark-quark and quark-diquark pairs. Classically, the longitudinal component of the transverse momentum of the colliding parton satisfies a sharp classification boundary between the energy independent and energy dependent generators as far as the elastic and inelastic collisions are concerned in the light of  SIBYLL 2.1 Ahn {\em et al.}(2009)  

\subsection{QCD-improved parton model} 

Importantly, the cosmic ray interaction event generator plays an important role Hiroaki {\em et al.}(2017) In this case, we aim to study stability of an interacting system with multiple soft interactions via the fluctuations of the gluon density in the limit of steeply rising double leading logarithmic approximation. We provide saddle point analysis of soft/ hard interactions and analyze the undermining statistical phase transitions, see Aurenche {\em et al.}(1992). for the data concerning the centre of mass of transverse component. Based on the dual parton model Aurenche {\em et al.}(1992) we classify stability property of hadronic interactions via proton-proton collision with a fixed energy scale as well as an energy dependent scale satisfying geometric saturation condition  Arringtin {\em et al.}(2007) in terms of the strong coupling constant 
\begin{equation}
\frac{\alpha_s(p_T^2)}{p_T^2}\ x g(x, p_T^2) \le \pi R_p^2,
\end{equation} 
where $R_p$ is the effective radius of the proton in the transverse space, $p_T$ is the transverse momentum and $x$ is the fractional energy of quarks in DPM picture. Physically, our motivations of the hadronic matter stability analysis follow from the elastic and inelastic cross sections for $p$-$p$ and $p$-$\overline{p}$ interactions Hiroaki {\em et al.}(2017). Hereby, the hard interaction happens at a point that correspond to approximately corresponds to a position where we have a constant value of the transverse momentum of the QCD improved partons.

Given the above qualification of soft interactions, we have a soft profile function corresponding to energy dependent minijet production via steeply rising gluon density, whereby we explore its fluctuations  in the space of the parameters $L$ and $Q$. As in Hiroaki {\em et al.}(2017). the hadron-hadron collision in the limit of double leading approximation going beyond the collinear factorization approximation are expected to correspond to a Gaussian distribution as a function of $\{L,Q \}$ which is the effective radius of a proton in the transverse space. Namely, for a proton-proton collision Hiroaki {\em et al.}(2017) mediated via the above geometric saturation condition in the limit of minijet model, with the minimum transverse momentum $q$, the corresponding gluon density of soft interactions Ahn {\em et al.}(2009) is given by 

\begin{equation} \label{gdo}
x g(x, q^2) \sim \exp \bigg[ \frac{48}{11-\frac{2}{3n_f}} 
\log\frac{\log\frac{q^2}{L^2}}{\log\frac{Q^2}{L^2}}
\log \frac{1}{x} \bigg]^{1/2}
\end{equation}

Subsequently with the above limiting gluon density function \ref{gdo}, we find its maxima/ minima of in the space of its parameters $\{L, Q\}$  at a fixed $b$ fractionation energy and effective radius of the proton in the transverse momentum with its dependence on the Bjorken scaling $Q^2 = p_T^2$ and the centre of mass energy square $s$. Herewith, we study the fluctuations of the gluon density concerning the stability of proton or other baryons, see Andreas {\em et al.}(2010).and A.S. Kronfeld Nora {\em et al.}(2004) provides a comprehensive report on quantum chromodynamics and associated heavy quark physics. One may further consider optimization of multiparticle profile in the light of two-component dual parton model, see  Aurenche {\em et al.}(1992) 

Let us focus on the pp collisions with its corresponding profile function as a vector valued function in the $st$-plane. Thus, the respective optimization of the boundary of soft intersection is obtained via the maxima/minima of the associated gluon density function in the space of its parameter $L$ and $Q$. Associated physical properties arises via the Bremsstrahlung, see Claude {\em et al.}(2012). In the light of classically field theory, variations of the action about its fixed points yields Euler Langrage equations. Here, the variations are performed with respect to all possible fields about fixed points of the undermining action integral of the theory, see in text in QFT Claude {\em et al.}(2012). chapter 2 and also in QFT Ashok {\em et al.} (2008) Furthermore, our fluctuation theory based study can incorporate effects of the optimal soft magnetic field, as well.

\subsection{Stability of fluctuating models}
Given the gluon density function $f(Q,L)$, its fluctuations are described through the embedding function reading as 

\begin{equation} \label{fql}
f:\mathcal{M}_2 \rightarrow \mathbb{R},
\end{equation}

In the above case of two fluctuation variables $(Q,L)$ with the above embedding $f(Q,L)$, it follows (Ruppeiner {\em et al.} (1995), Tiwari {\em et al.} (2011), Bellucci {\em et al.} (2010), Tiwari {\em et al.} (2011), Bellucci {\em et al.} (2013), Bellucci {\em et al.} (2013), Tiwari {\em et al.} (2018), Bellucci {\em et al.} (2018), Aman {\em et al.} (2006), Bellucci {\em et al.} (2010), Tiwari {\em et al.} (2011), (Weinhold\ Frank {em et al.} 1975); Ruppeiner {\em et al.} (1983), Ruppeiner {\em et al.} (1979), Ruppeiner {\em et al.} (1983), Millan {\em et al.} (2010), Simeoni {\em et al.} (2010), Widom {\em et al.} (1972)) that the correlation discontinuities give the nature of undermining phase transitions on the fluctuation surface $\mathcal{M}_2$. This arises when the system of the fluctuation matrix as in Eqn.(\ref{hm}) vanishes as a function of the QCD parameters $\{ Q,L \}$.

Herewith, from Eqn.(\ref{fql}), the stability analysis is carried around its critical points arising as the roots of the flow equation $ f_Q=0$ and $f_L =0$ by jointly evaluating the signs of one of the heat capacities $\{f_{QQ}, f_{LL} \}$ and the fluctuation determinant $\Delta: = |H|$ as the first and second principal minors of the concerning fluctuation matrix 
\begin{equation} \label{hm}
H:= \frac{\partial^2}{\partial x^i \partial x^j}f(Q,L)
\end{equation}
for the chosen parameter vector $\vec{x}:= (Q, L)$ on the surface $\Sigma $ of parameters $\{ Q, L\}$. 
In the light of statistical mechanics, we wish to classify the nature of an ensemble of QCD-improved Pomeron configurations with their gluon densities as in Eqn.(\ref{gdo}) having values of the parameters as
\begin{equation}
\Sigma := \{ (Q^{(1)}, L^{(1)}), (Q^{(2)}, L^{(2)}), (Q^{(3)}, L^{(3)}), \ldots, (Q^{(N)}, L^{(N)}) \}
\end{equation}
where it corresponds to a (non)interacting statistical basis as we take $N \rightarrow \infty$.

In the above set up, it is worth mentioning that we have a stable configuration when both the first principal minor $p_1: = f_{QQ}$ and the second principal minor $p_2:= \Delta$ take a positive value as defined in the next section. Notice further that the undermining system corresponds to an unstable statistical system when $p_2$ is positive and $p_1$ is negative. It is worth mentioning that the system corresponds to a saddle point configuration when we have a negative value of $p_2$ for any value of $p_1$, under fluctuations of the parameters $\{ Q, L\}$.

\subsection{Fluctuations in collision data}
Now we determine the optimum value of the transverse momentum which would correspond to soft profile function of proton. To be precise, this arises as a type of Taylor series expansion of the gluon density as in Eqn.(\ref{gdo}) with respect to the parameters ${Q, L}$. It is worth mentioning that the centre of mass component corresponding to the undermining pp collisions satisfies the Regge trajectory Hiroaki {\em et al.}(2017). Statistically, the parameters ${Q,L}$ give the variance of the gluon parametric density fluctuations that in the case of the collision data corresponds to a fuzzy area of approximate radius of the proton in transverse space.

The interaction area can be parametrized as a Gaussian fluctuation profile with an energy dependent width. Following the fundamental central limit theorem of statistics Huang {\em et al.}(2009 ) it follows that every fluctuating sample leads to a Gaussian profile in long term limit of interactions. In this concern, Ref. (Lucien {\em et al.}1986, Dmitry{\em et al.}2017)provides the proof of the classical central limit theorem and its variants. Physically, the constants such as the impact parameter play an important role in understanding the statistical behavior of a chosen pair of protons undergoing the collisions.

\section{Gluon density fluctuations}

In this section, following the fact that fixed target and collider experiments offer valuable guidance in modelling hadronic interactions Hiroaki {\em et al.}(2017), we provide fluctuation theory analysis of gluon density function. Note that in the setup of Lund string fragmentation model, if the quark has a stable distribution then the diquark will have an unstable distribution in the limit of DPM picture. 

Physically, fluctuations of an ensemble of gluons will continue in a damp manner untill the remaining mass of the string becomes smaller then the threshold mass of the quark/diquark pair mass. In other words the fluctuations stop when two final hadrons are formed. To discuss the above phenomena, one typically includes high multiplicity, increase in the mean transverse momentum, high transverse momentum jet and increase in the central rapidity density. The energy fractination plays an important role in high energy Lipatov {\em et al.}(1997) whereby the minijet model based on QCD improved parton configuration is expected to have an energy dependent transverse momentum cutoff Arringtin {\em et al.}(2007). In the limit of double leading approximation, we have steeply rising gluon density profile as in Eqn.(\ref{gdo}).
 
Further we study fluctuations of the gluon density as in Eqn.(\ref{gd}) as a function of the initial transverse momentum $Q$ and QCD renormalization scale $L$. In order to do so, we may examine the corresponding fluctuations of the gluon density Ahn {\em et al.}(2009) in the space of $\{ Q, L \}$ as the following  function  
\begin{equation} \label{gd}
f(Q, L) = B\log\frac{\log\frac{q^2}{L^2}}{\log\frac{Q^2}{L^2}},
\end{equation}
where the coefficient $B$ is related to the gluon density $g(x)$ at a fraction of energy $x$ and number of flavours $n_f$ by the relation
\begin{equation} \label{bnx}
B = \exp \bigg( -(\log(\frac{xg}{k}))^2 \frac{1}{\log \frac{1}{x}} \frac{11-\frac{2}{3}n_f }{48} \bigg)
\end{equation}

\subsection{Local fluctuations}
To study fluctuations of the above gluon density, we need to compute the corresponding flow components of gluon density $f(Q, L)$. In this case, see that the $Q$-flow component can be written as
\begin{equation}
\frac{df}{dQ} = -\frac{2B}{Q} \frac{\log\frac{L^2}{q^2}}{(\log \frac{Q^2}{L^2})^2}
\end{equation}
Similarly, it follows that the flow component corresponding to QCD renormalization scale $L$ reads as
\begin{equation} 
\frac{df}{dL}= \frac{2B}{L}\frac{\log\frac{q^2}{Q^2}}{(\log \frac{Q^2}{L^2})^2}
\end{equation}
Hereby, the flow equations $f_Q=0$ and $f_L=0$ yield the value of $L$ and $Q$ as $L=\pm q$ and $Q= \pm q$ respectively. Thus, it would be interesting to examine the nature of statistical interactions near an equal value of the QCD parameters $Q$ and $L$. To examine the nature of the stability of the gluon density profile as in Eqn.(\ref{gd}), we need to compute the fluctuation capacities defined as its second pure derivatives with respect to the system parameters $\{ Q, L \}$. 
In this case, we see that the pure $Q$-capacity of the gluon density simplifies as
\begin{equation}
\frac{d^2f}{dQ^2} = \frac{2B}{Q^2} \frac{\log\frac{q^2}{L^2}}{(\log \frac{Q^2}{L^2})^{3}}\big(4+\log\frac{Q^2}{L^2}\big)
\end{equation}
Therefore, at an equal value of the model parameters $Q=L$, we find that the fluctuation capacity $f_{QQ}$ diverges to a large positive value. Similarly, the corresponding pure $L$-capacity is given as per following expression
\begin{equation}
\frac{d^2f}{dL^2} = \frac{2B}{L^2}
\frac{\log\frac{q^2}{Q^2}}{(\log \frac{Q^2}{L^2})^3} \big(4-\log\frac{Q^2}{L^2}\big)
\end{equation}
In the above case as well, at an equal value of the model parameters $Q=L$, the fluctuation capacity $f_{LL}$ diverges to a large positive value. 
\subsection{Local correlation}
The undermining local correlation following the above gluon density as depicted in Eqn.(\ref{gd}) is given by its corresponding mixed derivative $f_{QL}$ as following 
\begin{equation}
\frac{d^2f}{dQdL} = - \frac{4B}{QL} \frac{\log \frac{q^4}{Q^2L^2}}{( \log \frac{Q^2}{L^2})^{3}}
\end{equation}
Thus, at the value of the model parameters $Q=q$, for all values of $q, L \in \mathbb{R}$, the local cross correlation $f_{QL}$ is given as 
\begin{equation}
\frac{d^2f}{dQdL}|_{Q=L} = - \frac{4B}{q^2} ( \log \frac{q^2}{L^2})^{-2}
\end{equation}
Later we study the equations at the critical values of the model parameters $Q=q$ and $L=q$ to find that the local cross correlation $f_{QL}$ diverges to a large negative value. 

\subsection{Global fluctuations}
In order to examine the global stability of the ensemble under fluctuations of $\{Q, L\}$, we are required to compute the fluctuation discriminant $\Delta:= f_{QQ} f_{LL} - (f_{QL})^2$. Substituting the above expressions for fluctuation capacities 
$\{ f_{QQ}, f_{LL}\}$ and cross correlation $f_{QL}$, we find the following fluctuation discriminant:
\begin{equation} \label{disc1}
\Delta(Q, L) = - \frac{4B^2}{Q^2L^2(\log(\frac{Q^2}{L^2}))^4}\bigg(4+\log(\frac{q^2}{Q^2})\ \log(\frac{q^2}{L^2}\bigg)
\end{equation}
Finally, at the value of $Q=q$, the determinant of the fluctuation matrix modulates as 
\begin{equation} \label{disc}
\Delta(Q,L) = - \frac{16 B^2}{Q^2L^2 \bigg(\log(\frac{Q^2}{L^2})\bigg)^4}
\end{equation}
We note that at the critical values of the model parameters $Q=q$ and $L=q$, the discriminant $\Delta$ undermining the global correlation diverges to a large negative value. Thus, as we approach the parametric fixed point $(Q,L) = (|q|, |q|)$, we have an unstable configuration as both the fluctuation capacity $\{ f_{ii}\ |\ i = Q, L \}$ and fluctuation determinant diverge.

\subsection{Physical Perspectives}

Note that when one of the parameter critical points $Q^2=q^2$ is approached while the QCD renormalization scale $L$ remains fixed, we have a locally stable configuration as the respective heat capacities $f_{QQ}$ or $f_{LL}$ take a positive value. However, even in the limit of approaching the transverse momentum square $q = \pm Q$ while $L^2$ remains fixed, the system remains unstable as the determinant $\Delta(Q, L$ diverges.

In the light of QCD phase transitions, we see that the correlation length does not diverge unless $ \Delta=0 $. That is, the point $\{Q, L\}$ in the parameter space $\mathcal{M}_2$ satisfies
\begin{equation} 
4 +\log \frac{q^2}{Q^2} \log \frac{q^2}{L^2}= 0
\end{equation}
Further, we have zero correlation length when the fluctuation determinant $ \Delta \rightarrow \infty $. This happens also when we either have $|Q|=\infty$ or $|L|=\infty$ or the function $B(n_f, x)= 0$ as in Eqn(\ref{bnx}). This would never happen in this fluctuation theory model of Pomerons because the function $f(Q, L)$ vanishes identically in this limit.

On the other hand, there are no statistical correlations when the correlation area $l^2 \sim \Delta^{-2} =0$. This happens for the following limiting values of the model parameters: $\{ Q, L \}$ as $Q=0$ or $L=0$ or $Q= \pm L$, or when the point $\{Q, L\}$ in the space of model parameters satisfying
\begin{equation} \label{deo}
4+\log(\frac{q^2}{Q^2})\ \log(\frac{q^2}{L^2})\rightarrow \infty
\end{equation}
It is evident that the Eqn.(\ref{deo}) holds whenever we have either $q>>L$ or $q>>Q$. Later we discuss the statistical interpretations the above results concerning the local and global correlations undermining an ensemble of QCD-improved Pomeron configurations where the associated gluon density is considered as a function of the parameters $\{Q, L\}$.

\subsection{Statistical Interpretations}

\begin{itemize}
\item In the light of the above analysis, the corresponding system is expected to yield a locally stable statistical basis when either of the heat capacities $\{ f_{QQ}, f_{LL} \}$ has a positive sign. For $q>>L$, the capacity $f_{QQ}$ happens to be positive when the factors $\log\frac{Q^2}{L^2}$ and $(4+\log\frac{Q^2}{L^2})$ have the same sign. 
That is, for $Q^2>L^2$, we have the following condition $Q^2>L^2/e^4$. On the other hand, for $Q^2<L^2$, the fluctuation capacity $f_{QQ}$ take a positive value for $Q^2<L^2/e^4$. 

Therefore, the local instabilities are expected to be detected in the region $L^2/e^4<Q^2<L^2$. Hereby, the test of stability needs further examinations in this band of parameters $\{Q, L\}$ as well as at their extreme values $Q=\pm L$, $q=\pm L$, and $q=\pm Q$. Similarly, if we choose the QCD renormalization scale $L$ as the  first coordinate in the surface of the parameters as $\{(L, Q) | L, Q \in \mathbb{R} \}$, for the choice of the transverse momentum $q>Q$, the capacity $f_{LL}$ happens to be positive when the factors $\log\frac{Q^2}{L^2}$ and $(4-\log\frac{Q^2}{L^2})$ have the same sign. The analysis follows as above in the case of $f_{QQ}$. 
\item The corresponding local correlation is defined as the mixed derivative of $f(Q, L)$ that reads as the ratio
\begin{equation}
f_{QL} = -\frac{4B}{QL}\frac{\log\frac{q^4}{Q^2L^2}}{(\log\frac{Q^2}{L^2})^3}
\end{equation}
For $Q>0$ and $L>0$, note the local correlation $f_{QL}$ as defined above remains negative for either $Q^2L^2>q^4$ and $Q>L$ or  $Q^2L^2<q^4$ since $q^2>> L^2$ since $q^2>>L^2$. Physically, in this case, their exists a locally contracting statistical system. On the other hand, for$ Q>0, L>0$, we have a locally expanding system when $f_{QL}$ takes a negative value, that is, we have either $Q^2L^2<q^4$ and $Q>L$, or $Q^2L^2>q^4$ and $Q<L$.
\item In this case, the phase transitions occur when the global correlation length $l \rightarrow \infty$. From Eqn.(\ref{disc}), we see that this happens precisely as per the correlation area 
\begin{equation}
l^2 \sim - \frac{1}{4B^2}\frac{Q^2L^2(\log\frac{Q^2}{L^2})^4}{(4+\log\frac{q^2}{L^2}\log\frac{q^2}{Q^2})}
\end{equation}
\item For the choice of $Q= L$, from the Eqn.(\ref{disc}) it follows that $\Delta(Q, Q) \rightarrow \infty$. Thus, the undermining correlation length $ l \sim \Delta^{-1} \rightarrow 0$. Physically, in the light of statistical mechanics, we may interpret that an ensemble of QCD-improved Pomeron configurations with their gluon densities having an ensemble of values $\{ (Q^{(i)}, L^{(i)})\ |\ i=1, 2, \ldots \infty \}$ of the QCD parameters $\{ Q, L\}$ correspond to a non interacting statistical configuration.
\end{itemize}

\section{Discussion of the results}

In this section, we give an interpretation of the results and offer their qualitative analysis in the light of existing literature.\\

Below, in the fig (1) we offer a qualitative behavior of the minimum transverse momentum following the geometric saturation condition Hiroak {\em et al.}(2017). For different value of the QCD renormalization scale, we see that it has a varied range of behavior namely for $L= 0.0065$, $Q$ has a similar behavior as SIBYLL 1.7 that uses constant parton density parameterization with energy independent transverse momentum cut off for proton-proton and proton-antiproton scattering cross section. For a higher value $L=0.065$, we see a larger flow of $Q$ plotted with respect to the energy in logarithmic scale with its discrete values as the following definition:
\begin{equation}
lnE_i^{(j)}=\frac{1}{2c^2} \bigg(\log(\frac{q_i}{L_j}) (1-\exp(-\frac{1}{ \log(\frac{q_i}{L_j})})\bigg)^2,\ j= 1, 2, 3
\end{equation}

\begin{center}
	\begin{figure}
		\hspace*{2.0cm} \vspace*{-0.4cm}
		\includegraphics[width=9.0cm,angle=0]{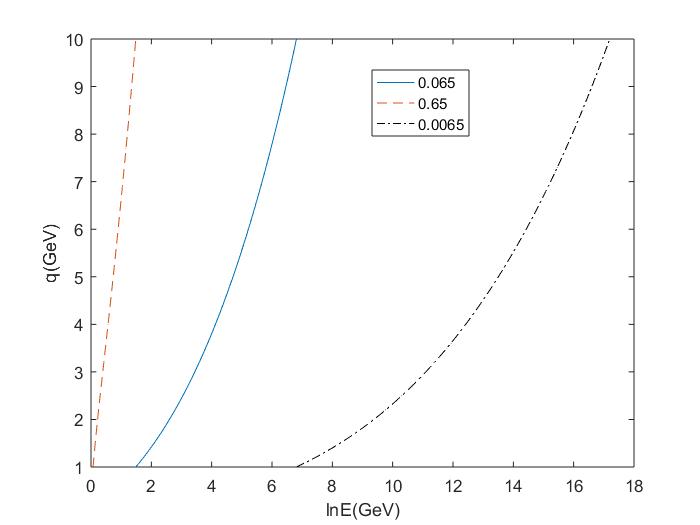}
\caption{The initial momentum $q$ plotted as a function of the energy $E$ in the logarithmic scale plotted on $X$-axis and $q$ on $Y$-axis describing the nature of flow of the momentum by considering variations of the energies $E$ for different values of $L$. Here, $L_1 = 0.065$ is depicted by the red solid line, $L_2 = 0.65$ by the blue dashed line and $L_3= 0.0065$ by the black dash-dot line.} \label{Gluon3}
		\vspace*{-0.01cm}
	\end{figure}
\end{center}

Furthermore, for an increase value of the QCD renormalization scale $L= 0.65$, the curve nearly becomes a straight line originating from the origin of the space of transverse momentum and energy in the logarithmic scale. Hereby, from fig.(\ref{Gluon3}), we see that our fluctuation theory analysis is indispensable for a larger value of renormalization scale $L$. In other words, the SIBYLL 1.7 is acceptable for a smaller QCD renormalization scale which is physically not the case in order to have asymptotic freedom of the partons.\\


In contrast to the minijet based SIBYLL 2.1, viz. the figure [1] of Hiroak{\em et al.}(2017), where the minimum transverse momentum is plotted with respect to the center of the mass energy, the minimum transverse momentum has a vanishing energy cutoff value of 1.5 GeV and reaches the SIBYLL 1.7 cut-off at 2.2 GeV. Here we note that the respective values of the minimum transverse momentum values of the zero, $1.7$ and $7$ in  GeV scale of the energy $E$. For a nonzero value of the energy, we see a parabolic behavior as the value of $E$ increases in the logarithmic scale.

In this concern, a comparative analysis of the flow of the transverse momentum $Q$ with respect to energy $E$ is summarized in the Table~\ref{tab1}.\\
\begin{table}
\centering
\par
\begin{tabular}{|c|c|c | c | c | c |c}
  \hline
 E(GeV)           &        1      &        10     &       30           &             53    &    200           \\
  \hline
Q(GeV)  &   1.24     &   -13.12     &      -25.41     &    -33.88       &       -60.77         \\
\hline
E(GeV)  &  210     &    630     &       1000  &     1800 &           \\
\hline
Q(GeV) &   -61.98      &  -94.30      &   -111.16   &    -135.87 &            \\
   \hline
  \end{tabular}
  \caption {A comparative analysis of the flow of the transverse momentum $Q$ with respect to energy $E$ that are measures in GeV.} 
  \label{tab1}
\end{table}
\\
\\
 \vspace*{-1cm}
\begin{center}
	\begin{figure}
		\hspace*{2.0cm} \vspace*{-0.5cm}
		\includegraphics[width=8cm,angle=0]{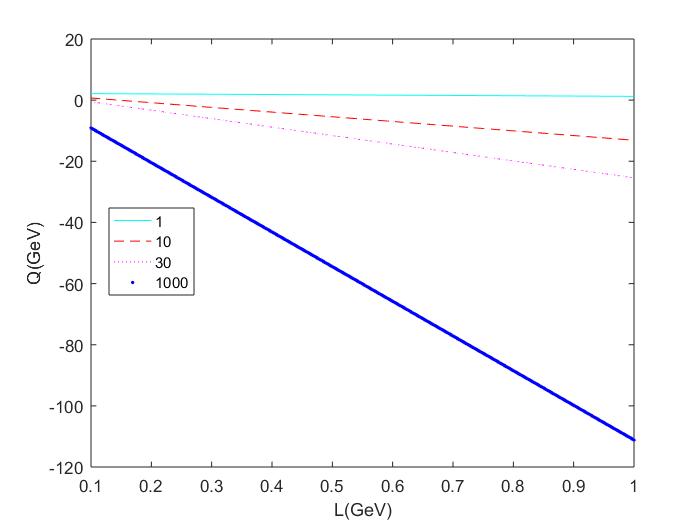}
		\caption{The initial momentum $Q$ plotted as a function of the energy $E$ in the logarithmic scale plotted on $X$-axis and $Q$ on $Y$-axis describing the nature of the momentum flow by considering variations of the QCD re normalization scale $L$ for different values of $E$. Here, the string fragmentation with the energy $E =1\ MeV$ is depicted by the cyan solid line, the diffraction dissociation pomeron with energy scale $E = 10\ MeV$ by the red dashed line, minijet production with energy scale $E= 30\ MeV$ by the magenta dotted line and the soft interactions with energy scale $E= 10000\ MeV$ by the blue point line.} \label{Gluon5}
		\vspace*{-0.5cm}
	\end{figure}
\end{center}

\begin{center}
	\begin{figure}
		\hspace*{2.5cm} \vspace*{-0.5cm}
		\includegraphics[width=8cm,angle=0]{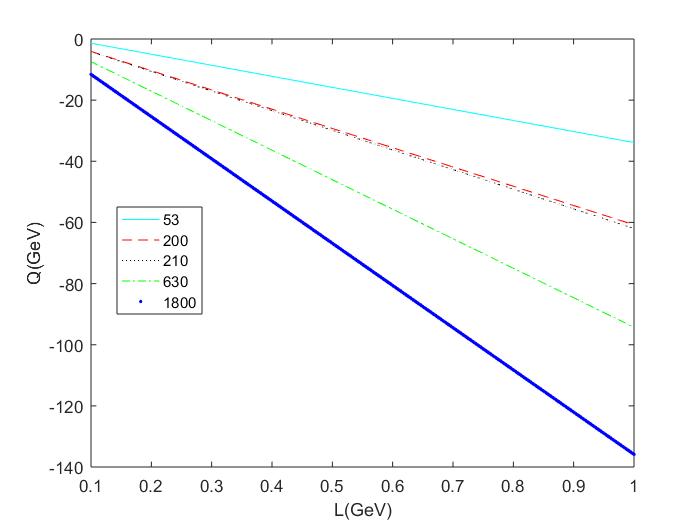}
    		\caption{The initial momentum $Q$ plotted as a function of the energy $E$ in the logarithmic scale plotted on $X$-axis and $q$ on $Y$-axis describing the nature of the momentum flow by considering variations of the energies $E$ for different values of $L$ for various collaborations. Here, the UAS collaboration $E= 53 MeV$ is depicted by the cyan solid line and its next energy $E = 200 MeV$ by the red dashed, GEUS collaboration energy scale $E= 210 MeV$ by the black dotted line, P238 collaboration with its energy scale $E= 630 MeV$ by the green dash-dot line, and CDF collaboration with its energy scale $E= 18000 MeV$ by the blue point line.} \label{Gluon6}
		\vspace*{-0.5cm}
	\end{figure}
\end{center}

In the fig.(\ref{Gluon5}), we provide qualitative discussion of diffraction dissociation based on the pomerons, soft interaction, string fragmentation and particle produced in minijet model with the definition of the energy as

\begin{equation}
Q(i,j)=q- L(j)\exp(c \sqrt{2 \log(E(i))}),
\end{equation}

where $E$ reads as above the caption of  fig.(\ref{Gluon5}, \ref{Gluon6}) and $L \in (0, 1)$. Namely, from fig.(\ref{Gluon5}), we find that the string fragmentation arises with a highest transfer momentum $q$ and soft interaction arise with a lowest transfer momentum $q$ when they are varied with respect to QCD renormalization scale $L$. In this analysis, we observe that the pomeron have a larger transverse momentum cutoff in comparison to the minijet model based particle productions. In all the above cases, we notice that the linear momentum cutoff almost behaved linearly with respect to the QCD renormalization scale.\\
 \vspace*{-0.5cm}

In the fig.(\ref{Gluon6}), we offer a qualitative analysis of the transverse momentum cutoff when it is varied with respect to  the QCD renormalization scale in order to compare with  the existing QCD collaboration energy scale Hiroaki{\em et al.}(2017). Hereby, from  fig.(\ref{Gluon6}), we see that UAS energy scale has a larger transverse momentum cutoff in comparison with the GEUS, P238 and CDF collaboration scale a comparative analysis of the transverse momentum cutoff for the energy scale of these collaboration shows that Q generally takes a negative value which physically indicates that the undermining particle is produced by absorbing a finite amount of energy. Namely, from  fig.(\ref{Gluon6}), we notice that the absorption energy increase as we increase QCD renormalization scale.

In the fig.(\ref{Gluon5}) and  fig.(\ref{Gluon6}), a comparative analysis of QCD transverse momentum cutoff with respect to the energy is summarized in the following table. Hereby, based on the Bjorken scaling in QCD cutoff, we have classified the nature of the transverse momentum, i.e., the energy emitted or absorb in due course of the gluon density fluctuation.
 \vspace*{-0.5cm}
\begin{center}
	\begin{figure}
		\hspace*{1.5cm} \vspace*{-0.6cm}
		\includegraphics[width=10.0cm,angle=0]{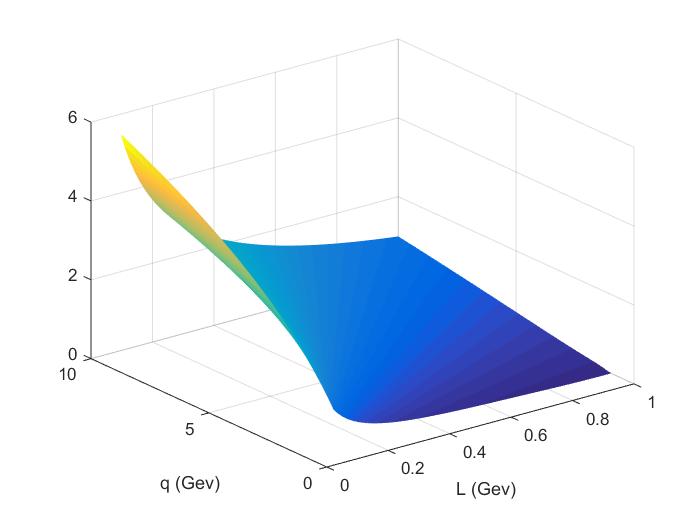}
		\caption{The emitted or absorb energy $E$ plotted in the logarithmic scale as a function of the transverse momentum $q$ on $Y$-axis and $L$ on $X$-axis describing the nature of the energy fluctuations by considering simultaneous variations over $q$ and $L$.} \label{Gluon7}
		\vspace*{-0.5cm}
	\end{figure}
\end{center}

Following the above observation as in fig.(\ref{Gluon5}, \ref{Gluon6}), we provide the qualitative analysis of the energy emitted or absorb during the formation of parton in the space of QCD transverse momentum cutoff and QCD renormalization scale in fig.(\ref{Gluon7}) with the energy definition as

\begin{equation}
\ln E_{i, j}=\frac{1}{2c^2} \bigg( \log(\frac{q_i}{L_j}) \big(1-\exp(-\frac{1}{\log (\frac{q_i}{L_j})})\big) \bigg)^2, 
\end{equation}

where $q_i \in (0, 10)$ and $L \in (0, 1)$. Here, we find that most of the emissions happen for a large QCD transverse momentum and small QCD renormalization scales. In other words, we find that there is almost no energy change for a small transverse momentum and large QCD renormalization scale. The corresponding behavior as shown in fig.(\ref{Gluon7}), where the energy augments to the order of a fixed value in its logarithmic scale as the transverse momentum takes a large value of the order $10$ for a given small QCD renormalization scale $L$. It is anticipated that the phase transition is expected to arise near a small value of the transverse momentum and QCD renormalization scale. Such processes are truly quantum in their nature that we leave it open for a future research investigation.
 \vspace*{-0.5cm}
\begin{center}
	\begin{figure}
		\hspace*{1.5cm} \vspace*{-0.6cm}
	   \includegraphics[width=10cm]{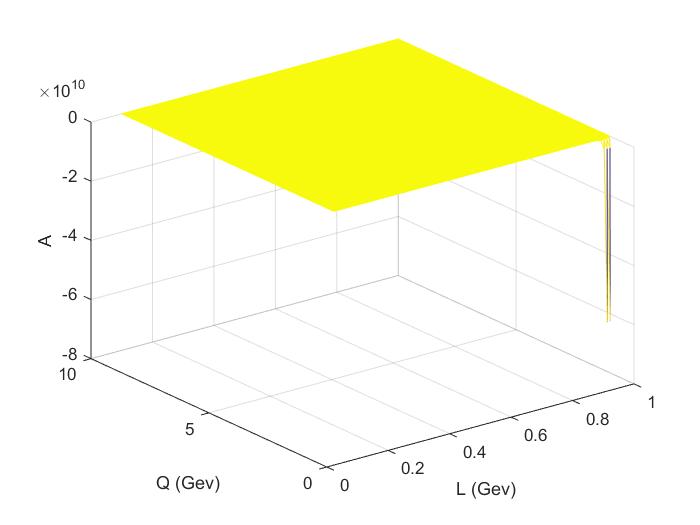}
		\caption{The discriminant $A:= \Delta$ plotted as a function of the transverse momentum $q$ on $Y$-axis and $L$ on $X$-axis describing the nature of the energy fluctuations by considering simultaneous variations over $q$ and $L$.} \label{Gluon10}
		\vspace*{-0.5cm}
	\end{figure}
\end{center}

In fig.(\ref{Gluon10}) with $x=0.1$, $g=10$, $k=1$, $n_f=3$, we provide a qualitative  behavior of the gluon density fluctuations in the space of the initial transverse momentum $Q$ and QCD renormalization scale $L$. Here for $B=25$, the discriminant $\Delta$ takes a large negative value of the order $8 \times 10 ^{10}$. We notice that the value of the amplitude depends on the choice of the prefactor $B$. From the functional definition of the fluctuation discriminant 

\begin{equation}
\Delta (i,j) = -\frac{16B^2}{q(i)^2 l(j)^2}
\bigg( \log \big(\frac{q(i)^2}{l(j)^2} \big) \bigg)^{-4},
\end{equation}

we observe that the sample becomes ill-defined when the initial transverse momentum, the minimum transverse momentum $Q$ and QCD renormalization scale $L$ take an identical value. 
In this case, from fig.(\ref{Gluon10}), we see that the fluctuation theory analysis needs to be prolonged with a higher scale to check the stability of the configurations under fluctuations of the gluon density in the space of $\{Q, L\}$ . Such studies we leave open for a future research.

\section{Conclusion and future directions}
In this paper, we examine an extrapolated equivalent description of gluon density fluctuations with respect to the initial transverse momentum and QCD renormalization scale fluctuate in the region where the proton has an effective radius in the transverse momentum space. This offers an intrinsic statistical understanding of overall structure of soft interaction and diffraction dissociation phenomena. We also addressed some of the shortcomings of the SIBYLL 2.1 by considering an ensamble of QCD improve parton models and analysing beyond the nucleus-nucleus collisions using semisuperposition model and the full Glauber model. Our analysis can as well be applied to  understand the  antibaryon production,namely, the distribution of anti protons in the central region of collisions Alexopoulos {\em et al.}(1996).

In this concern, we anticipate exploring the dependance of impact parameter of the collision governing the relevant gluon density of the target nucleus Alexopoulos {\em et al.}(1996).
 Such an understanding following our statistical analysis could improve the modelling of minijet model as a constraint minijet model with a refined profile function, whose analysis is relegated for a later publication. Note that our analysis entails the properties of hadron-nucleus and nucleus-nucleus interactions as achieved in Glauber setting Glauber {\em et al.}(1970). using two channel model, whereby we offer statistical properties of proton-proton interactions. 

It would be interesting to examine the fluctuation theory based understanding of rapidity and Feynmann distribution as in NA49 experiments Gornaya {\em et al.}(2018).and pseudorapidity and multiplicity distribution of charged particles in central region. Indeed, our model by no means gives the final answer and most likely exhibits short comings in the limits where the statistical approximation breaks down. 

With the above advantages and disadvantages of the double leading logarithmic approximation of the gluon density function,we emphasize that our statistical analysis is able to successfully reproduce the increasing nature of transverse momentum as energy scale is varied in many of experimental results. This includes that energy scales of diffraction dissociation particles such as Pomerons, soft interaction models in the light of Regge theory, minijet productions,string fragmentation of quark-quark and quark-diquark pairs, CDF, P238, UAS, GEUSS,and UA4 collaborations (Nik {\em et al.}2018, Zha {\em et al.}2018, Szabolcs {\em et al.}2018, McNelis {\em et al.}2018, Khoze {\em et al.}2018). We further believe that application of our fluctuation analysis theory would offer an improved understanding  of other models in analysing the behaviour of the very high energy particles such as the analysis of extensive air shower  and cosmic rays.

The study of multiple interacting soft systems composed of the hadron-hadron collisions with  given profile functions play an important role. Namely, the proton-proton collision profile function is anticipated by using fuzzy Gaussian soft interactions with an exponential form factor corresponding to the Gaussian profile function of the constituent protons, see Bruyn {\em et al.}(2018). for monojet and trackless jets as the final states towards developments of the CMS Detector at the LHC. Such a model can be obtained as an effective configuration of the two body interaction system. Therefore, exploring the stability of gluonic matter in the soft/ hard limit as a function of the energy and associated physical meaning in the limit of vanishing impact parameter is in the future scope of this research. Namely, for a given data, we wish to experimentally characterize an optimal value for the transverse size of a proton at an energy scale $\sqrt{s}$ for fixed values of initial energy $s_0$ and Regge slope $\alpha^{\prime}$. In nutshell, because of multi-particle interactions, we anticipate that an energy dependent transverse momentum as function of the energy square $s$ and parameters of the chosen fluctuation dynamics as a random observation sample. Such an analysis lies in the future scope of this research.


\begin{thebibliography}{99}
\bibitem{6r} Alekhin,S.et al.2010,``3-, 4-, and 5-flavor next-to-next-to-leading order parton distribution functions from deep-inelastic-scattering data and at hadron colliders.” {\it Phys. Rev. D} 81, 014032, DOI: 10.1103/81.014032.

\bibitem{CDF}  Aminzadeh, N.R.,et al.2018, ``Study of inclusive single-jet production in the framework of k t-factorization unintegrated parton distributions.” {\it Phys. Rev. D} 97, no. 9, 096012.

\bibitem{2r} Amaral, J.T. et al.2014,``Probing gluon number fluctuation effects in future electron-hadron colliders.” {\it Nuc. Phys. A},930, 104, DOI:  10.1016/j.nuclphysa.2014.08.015.

\bibitem{kc} Andreas, K.et al.2010,  Resource Letter QCD-1: Quantum chromodynamics. {\it American Journal of Physics} 78, no. 11, 1081-1116.

\bibitem{23r} Arringtin, J.et al.2007, C. D. Roberts, and J.M. Zanotti, Quadrupole moment and a proton halo structure in $^{17}F (I_{\pi} = 5/2^+)$. {\it J. Phys. G} 34, 523.


\bibitem{12r} Ahn, E.J. et al.2009,Cosmic ray interaction event generator SIBYLL 2.1. {\it Physical Review D},80, 094003, DOI: 10.1103/.80.094003.

\bibitem{eegls} Ahn,E.J. et al.2009, Cosmic ray interaction event generator SIBYLL 2.1.{\it  Physical Review D} 80, no. 094003.

\bibitem{56} Alexopoulos, T. et al.1993, (E735 Collaboration), Mass-identified particle production in proton-antiproton collisions at $\sqrt{s} = 300, 540, 1000,\  \mbox{and}\ 1800${\it  Phys. Rev. D} 48, 984.

\bibitem{aman} Aman, et al.2006,"Flat information geometries in black hole thermodynamics." {\it General Relativity and Gravitation} 38, no. 8: 1305-1315. {\tt ArXiv:gr-qc/0601119v1}.

\bibitem{18b} Adloff, C. et al.1997, ``A Measurement of the Proton Structure Function” $F_2(x, Q^2)$ at Low $x$ and Low $Q^2$ at HERA, {\it Nucl. Phys. B} 497, 3.

\bibitem{cd} Alekhin, S. et al.2018,``Strange sea determination from collider data." {\it Physics Letters B} 777: 134-140.

\bibitem{0br}  Andronic, A.,et al.2018, ``Decoding the phase structure of QCD via particle production at high energy”. Nature, 561 (7723), 321, DOI: 10.1038/s41586-018-0491-6.

\bibitem{1br} Andronic, A.,et al.2018, ``Decoding the phase structure of QCD via particle production at high energy”.{\it Nature volume} 561, 321-330.

\bibitem{9r} Aaron, F.D. et al. 2010, (H1 Collaboration), ``Combined Measurement and QCD Analysis of the Inclusive ep Scattering Cross Sections at HERA” {\it J. High Energy Phys}, 01,109, DOI: 10.1007.
	

\bibitem{5r} Arsene, I. et. al. 2005, ``Quark Gluon Plasma and Color Glass Condensate at RHIC? The perspective from the BRAHMS experiment.{\it Nuclear Phys.A}
\bibitem{33r} Aurenche, A. et al. 1992,``Multiparticle production in a two-component dual parton model.” {\it Phys. Rev. D} 45, 92.


\bibitem{prdref} Aurenche, P.,et.al, 1992,``Multiparticle production in a two-component dual parton model.” {\it Phys. Rev. D} 45, no. 1, 92.


\bibitem{11br} Bellwied, R.,et al. 1994,“The QCD phase diagram from analytic continuation." {\it Physics Letters B} 751, 559-564, DOI: 10.1016/j.physletb.2015.11.011.

\bibitem{19b} Breitweg, J. et al. (ZEUS Collaboration), 2002. ``The $\gamma^{star}p$ total cross section and elastic diffraction." {\it Phys. Lett. B}407, 432.

\bibitem{3r} Berger, E.L. et al. 1992,``The gluon density.” ANL-HEP-CP-92-108; CERN-TH-6739/92;\\ CONF-921122-17 ON: DE93005565.\\{\it Phys Review D},80, 094003, DOI: 10.1103/.80.094003

\bibitem{ourpaper} Bellucci,S.,Tiwari,B.N. 2010, "State-space correlations and stabilities." {\it Physical Review D} 82, no. 8: 084008. 

\bibitem{6br}  Bass, S.A, et al. 1999,``Signatures of quark-gluon plasma formation in high energy heavy-ion collisions: a critical review." {\it Journal of Physics G}: Nuclear and Particle Physics 25, no. 3.

\bibitem{bookbtg} Bellucci, S. et al. 2013, ”Geometrical methods for power network analysis.” (SpringerBriefs in Electrical and Computer Engineering) ISBN: 978-3-642-33343-9.

\bibitem{18r} Bahr, M. et al.2007, "Herwig++ 2.1 Release Note".\\ arXiv:0711.3137.


\bibitem{bntngsb} Bellucci, S. et al.2013,``Intrinsic Geometric Characterization." In Geometrical Methods for Power Network Analysis”, pp. 19-28. Springer, Berlin, Heidelberg.

\bibitem{fisica} Bellucci, S. et al.2011,“Evolution, correlation and phase transition" Physica A: Statistical Mechanics and its Applications 390, no. 11: 2074-2086. e-print:{ArXiv:1010.5148v1 [stat-phys]}.

\bibitem{halfbpspaper} Bellucci, S.et al.2010,``An exact fluctuating 1/2-BPS configuration." {\it Journal of High Energy Physics} {\tt arXiv:0910.5314v2 [hep-th]}.

\bibitem{7r} Ball, R.D. et al.2010, (Neural Network PDF Collaboration), ``A first unbiased global NLO determination of parton distributions and their uncertainties." {\it Nucl. Phys B},838,136, DOI: 10.1016/j.nuclphysb.2010.05.008

\bibitem{Bruyn} Bruyn, D. 2018, “Search for Dark Matter in the Monojet and Trackless Jets Final States with the CMS Detector at the LHC”,  PhD diss., Vrije U., Brussels.

\bibitem{Capella} Capella, A.,et al.1994, "Dual parton model." {\it Physics Reports},236, no. 4-5: 225-329.

\bibitem{4br} Cheng,M.et al.2006, ``Transition temperature in QCD." {\it Physical Review D} 74, no. 5, 054507.
	

\bibitem{zuber} Claude,.I.,et al.2012,  ``Quantum field theory.” Courier Corporation.

\bibitem{Das}Das, A. et al.2008,``Lectures on quantum field theory.” World Scientific.

\bibitem{clt2} Dmitry, D. et al.2017,``On mixing and the local central limit theorem for hyperbolic flows." Ergodic Theory and Dynamical Systems”, 1-33.

\bibitem{20r1} Drescher,H.J.et al.2001,  ``Parton-Based Gribov-Regge Theory.”{\it  Phys. Rep}, 350, 93.

\bibitem{12b} Engel, J.et al.1992,``Nucleus-nucleus collisions and interpretation of cosmic-ray cascades”, {\it Phys. Rev. D }46, 5013.

\bibitem{13b} Engel,R.et al.1999,``Primary spectrum to 1 TeV and beyond" in proceedings of the $26^{th}$ International Cosmic Ray Conference, Salt Lake City, Utah",(AIP, Melville, NY,2000), Vol. 1, p. 415.

\bibitem{19r3} Engel, R.et al.2001,``Models of primary interactions." Proceedings of the 27th ICRC, Hamburg, (Copernicus, Gesellschaft), 1381.

\bibitem{22r1} Engel,R.et al.1992, ``Nucleus-nucleus collisions and interpretation of cosmic-ray cascades." {\it Phys. Rev. D} 46, 5013.

\bibitem{15b} Fletcher, R.S, et al.1994,``The gap survival probability and diffractive dissociation”, {\it Phys. Lett. B} 320, 373, DOI: 10.1016/0370-2693(94)90672-6.

\bibitem{22r2} Fletcher, R.S, et al.1994, ``SIBYLL - An event generator for simulation of high energy cosmic ray cascades.” {\it Phys. Rev. D} 50, 5710,DOI:  10.1103/PHYSREVD.80.094003.

\bibitem{Weinhold1}Frank,W. et al.1975,``Metric geometry of equilibrium thermodynamics” {\it Journal of Chemical Physics} 63, no. 6: 2479-2483,DOI:10.1063/1.431689.

\bibitem{Weinhold2} Frank, W.1975,``Metric geometry of equilibrium thermodynamics. II. Scaling, homogeneity, and generalized Gibbs–Duhem relations." {\it Chemical Physics} 63, no. 6: 2484-2487, doi/10.1063/1.431635.
\bibitem{11b} Glauber , R.J. et al.1970, ``High-energy scattering of protons by nuclei”,{\it Nucl. Phys. B}21,135.

\bibitem{20b} Gluck, M.et al.1995, ``Dynamical parton distributions of the proton and small-$x$ physics.” {\it Z. Phys. C}67, 433.

\bibitem{21b}Gluck, M.et al.1998, ``Dynamical Parton Distributions Revisited.” {\it Eur. Phys. J. C} 5, 461.

\bibitem{14b}Good,M.L.et al.1960, ``Diffraction Dissociation of Beam Particles”, {\it Phys. Rev}, 120-1857, 10.1103/PhysRev.120.1857.
	

\bibitem{4r} Gonçalves,V.P.et al.2016,L. A. S. Martins, W. K. Sauter,  ``Probing the gluon density of the proton in the exclusive photoproduction of vector mesons at the LHC: a phenomenological analysis.” {\it Eur. Phys. J. C} 76: 97.

\bibitem{NA} Gornaya,J.et al.2018, ``Elliptic Flow of $\pi$–Measured with the Event Plane Method in Pb-Pb Collisions at 40A Ge in the NA49 Experiment at the CERN SPS. KnE Energ." Phys.3, 441-446.

\bibitem{16r}Goncalves,V.P.et al.2002, “Peripheral heavy ion collisions as a probe of the nuclear gluon distribution”.{\it Physical Review C}, 65,054905.

\bibitem{SIBYLL} Menjo, H. et al.2017, ``Status of the LHCf experiment".{\it Proceedings of Science}, Volume 301 - 35th International Cosmic Ray Conference (ICRC2017) - High-light Talks, DOI:10.22323/1.301.1099.

\bibitem{15r} Heinz,U.2000,``The Little Bang: Searching for quark-gluon matter in relativistic heavy-ion collisions.”{\it Nucl. Phys. A}685,414, DOI: 10.1016/S0375-9474(01)00558-9.

\bibitem{1r} Heikki,M.2016,``antysaari et al. ``Revealing proton shape fluctuations with incoherent diffraction at high energy.” {\it Physical Review D} DOI: 10.1103/PhysRevD.94.034042.

\bibitem{statistics}Huang,K.2009, ``Introduction to statistical physics.” Chapman and Hall/CRC.

\bibitem{7br} He, Min, Wei-min Sun, Hong-tao Feng, and Hong-shi Zong.2007,"A model study of QCD phase transition." {\it Journal of Physics G}: Nuclear and Particle Physics 34, no. 12, 2655.

\bibitem{15r} Heinz,U.2000, “ The Little Bang: Searching for quark-gluon matter in relativistic heavy-ion collisions”.{\it Nucl. Phys. A}685,414.

\bibitem{Jain} Jain,P.et al.2015,``The fine tuning of the cosmological constant in a conformal model.” {\it International Journal of Modern Physics A} 30, no. 32, 1550171, DOI:  10.1142/S0217751X15501717.

\bibitem{111br} Kalashnikov,O.K.et al.1979, `Phase transition in the quark-gluon plasma."{it Physics Letters B} 88, no. 3-4, 328-330, DOI: 10.1016/0370-2693(79)90479-9.

\bibitem{21r1} Kalmykov,N.N.et al.1979, ``EPOS Model and Ultra High Energy Cosmic Rays." {\it Nucl. Phys. B}, Proc. Suppl. 52B, 17.

\bibitem{UA4} Khoze, V. A.et al.2018, “ Elastic and diffractive scattering at the LHC.” ArXiv preprint ArXiv:1806.05970.
\bibitem{kc} Kronfeld,et al.2010,Resource Letter QCD-1: Quantum chromodynamics. {\it American Journal of Physics} 78, no. 11 1081-1116.

\bibitem{10r} Khoze,V.A.et al. A. D. Martin, M. G. Ryskin, ``Elastic and diffractive scattering at the LHC." Report number: IPPP/18/42  ArXiv:1806.05970v2.
	

\bibitem{clt1} Lucien,Le.C,et al.1986,``The central limit theorem around 1935. Statistical science.” 78-91.

\bibitem{24b} Lipatov,L.N.et al.1997, Small-x physics in perturbative QCD, {\it Phys. Rep}, 286, 131.

\bibitem{3br} Meyer-Ortmanns.1996, ``Phase transitions in quantum chromodynamics."{\it Reviews of Modern Physics} 68, no. 2, 473,DOI: 10.1103/RevModPhys.68.473.

\bibitem{McMillan} Millan,M.et al.2010, ``Fluid phases: Going supercritical." {\it Nature Physics} 6, no. 7: 479-480.

\bibitem{mutta} Muta,T.2009, Foundations of Quantum Chromodynamics: An Introduction to Perturbative Methods in Gauge Theories, ISBN-13: 978-9812793546, ISBN-10: 9812793542.

\bibitem{3br} Meyer-Ortmanns, Hildegard.1996, "Phase transitions in quantum chromodynamics." {\it Reviews of Modern Physics}68, no. 2: 473.


\bibitem{11r} Martin,A.D.et al.2009, “Parton distributions for the LHC” {\it Eur. Phys. J. C} 63: 189-285.

\bibitem{McMillan} McMillan, et al.2010, "Fluid phases: Going supercritical." {\it Nature Physics} 6, no. 7 (2010): 479-480.

\bibitem{GEUSS}  McNelis,et al.2018, “(3+ 1)-dimensional anisotropic fluid dynamics with a lattice QCD equation of state”. {\it Physical Review C} 97, no. 5054912.

	\bibitem{CDF}  Nik,et al.2018, Study of inclusive single-jet production in the framework of k t-factorization unintegrated parton distributions. {\it Physical Review D} 97, no. 9, 096012.
	
	\bibitem{Kronfeld} Nora,B.et al.2004, ``Heavy quarkonium physics”. ArXiv preprint hep-ph/0412158

\bibitem{20r2} Ostapchenko,S.S.et al.2001,  ``Proceedings of the 12th International Symposium on Very High Energy Cosmic Ray Interactions." Proceedings of the 27th ICRC Hamburg,  (Copernicus, Gesellschaft,), p. 446.

\bibitem{8r} Pumplin,J.et al.2009, ``Exploring portals to a hidden sector through fixed targets.” {\it Phys. Rev. D} 80, 014019, DOI: 10.1103/PhysRevD.80.095024.

\bibitem{5br} Rapp,R.et al.2002, ``Chiral symmetry restoration and dileptons in relativistic heavy-ion collisions." {\it Advances in Nuclear Physics}, pp. 1-205.

\bibitem{19r1} Ranft,J.et al.1995, ``Dual parton model at cosmic ray energies.” {\it Phys. Rev. D} 51, 64.

\bibitem{19r2} Roesler,S.et al.2001, Proceedings of the International Conference on Advanced Monte Carlo for Radiation Physics, Particle Transport Simulation and Applications\\, Lisbon, Portugal, 2000, edited by A. Kling, F. Barao, M. Nakagawa, L. Tavora, P. Vax (Springer Verlag, Berlin); Proceedings of the 27th ICRC, Hamburg, (Copernicus, Gesellschaft), p. 439.

\bibitem{rupp1} Ruppeiner,G.et al.1995, ``Riemannian geometry in thermodynamic fluctuation theory." {\it Rev. Mod. Phys}, 67, no. 3 : 605, [Erratum, 313-313, 68,]

\bibitem{rupp2} Ruppeiner,G.et al.1983, ``Thermodynamic critical fluctuation theory." {\it Physical Review Letters}, 50, no. 5: 287.

\bibitem{Ruppeiner3} Ruppeiner,G.et al.1979, “Thermodynamics: A Riemannian geometric model." {\it Physical Review A} 20, no. 4: 1608.


\bibitem{Ruppeiner4} Ruppeiner,G.et al.1983, ``New thermodynamic fluctuation theory using path integrals." {\it Physical Review A} 27, no. 2: 1116.

\bibitem{5br} Rapp,et al.2002,"Chiral symmetry restoration and dileptons in relativistic heavy-ion collisions." {\it Advances in Nuclear Physics}, pp. 1-205. Springer, Boston.

\bibitem{19r1} Ranft,J,1995," Dual parton model at cosmic ray energies." {\it Phys. Rev. D} 51, 64.

\bibitem{19r2}Roesler,S.et al.2001, Proceedings of the International Conference on Advanced Monte Carlo for Radiation Physics, Particle Transport Simulation and Applications\\
Lisbon, Portugal.2000, edited by A. Kling, F. Barao, M. Nakagawa, L. Tavora, P. Vax (Springer Verlag, BerliN); Proceedings of the 27th ICRC, Hamburg.

\bibitem{rupp1} Ruppeiner, G.et al.1995, "Riemannian geometry in thermodynamic fluctuation theory." {\it Reviews of Modern Physics} 67, no. 3 : 605., [Erratum, 313-313, 68, 1996].

\bibitem{rupp2} Ruppeiner, G.et al.1983,"Thermodynamic critical fluctuation theory?." {\it Physical Review Letters}50, no. 5: 287.

\bibitem{Ruppeiner3} Ruppeiner, G.et al.1979, "Thermodynamics: A Riemannian geometric model." {\it Physical Review A} 20, no. 4: 1608.

\bibitem{Ruppeiner4} Ruppeiner, G.1983,"New thermodynamic fluctuation theory using path integrals." {\it Physical Review A} 27, no. 2: 1116.

\bibitem{2br} Stephanov,M.A,2006, ``QCD phase diagram: an overview." arXiv preprint hep-lat/0701002.

\bibitem{Hera} Sarkar,A.C,et al.2018, ``Impact of low-x resummation on QCD analysis of HERA data.” {\it Eur. Phys. J. C} 78: 621.


\bibitem{17r} Sjostrand,T.et al.2006, ``PYTHIA 6.4 physics and manual." {\it J. High Energy Phys},\\DOI: 10.1088/1126-6708/2006/05/026.

\bibitem{Simeoni} Simeoni,G.et al.2010, ``The Widom line as the crossover between liquid-like and gas-like behaviour in supercritical fluids." {\it Nature Physics} 6, no. 7: 503-507.

\bibitem{UAS} Szabolcs,B.et al.2018,``Higher order fluctuations and correlations of conserved charges from lattice QCD.” ArXiv preprint arXiv:1805.04445.

\bibitem{2br} Stephanov, M.A.2006, "QCD phase diagram: an overview." ArXiv preprint hep-lat/0701002.

\bibitem{17r} Sjostrand, T.et al.2006, {\it J. High Energy Phys}.

\bibitem{21r2} Sciutto,S.J.et al.2001, Proceedings of the 27th ICRC, Hamburg, (Copernicus, Gesellschaft,).

\bibitem{Simeoni} Simeoni,et al.2010, "The Widom line as the crossover between liquid-like and gas-like behaviour in supercritical fluids." {\it Nature Physics} 6, no. 7 : 503-507.

\bibitem{book2018} Tiwari,.B.N.et al.2018,  Randomized Cunningham Numbers in Cryptography: Randomization theory, Cryptanalysis, RSA cryptosystem, Primality testing, Cunningham numbers, Optimization theory (LAP LAMBERT Academic Publishing), ISBN-13: 978-6139858477.

\bibitem{tcb} Tiwari,B.N.et al.2011, A Thermodynamic Geometric Study Of Complex Entropies: Statistical Fluctuations: Shannon, Renyi, Tsallis, Abe And Structural Configurations, (LAP LAMBERT Academic Publishing,) ISBN-13: 978-3845420691.

\bibitem{bntfr} Tiwari,B.N.2011, Sur les corrections de la g\'{e}om\'{e}trie thermodynamique des trous noirs, (\'Editions Universitaires Europ\'eennes, ), ISBN: 978-613-1-53539-0 {\tt ArXiv:0801.4087v2 [hep-th]}.

\bibitem{bntef} Tiwari,B.N.2011, Geometric Perspective of Entropy Function: Embedding, Spectrum and Convexity, (LAP Lambert Academic Publishing), ISBN-13: 978-3-8454-3178-9 {\tt http:arxiv.org/abs/1108.4654v2 [hep-th]}.

\bibitem{widom} Widom,B.1972,``Surface Tension of Fluids, in Phase Transitions and Critical Phenomena.” Vol. 2 (eds C. Domb and J. L. Lebowitz), (Academic Press).

\bibitem{Weinhold1} Weinhold,F.1975, "Metric geometry of equilibrium thermodynamics." {\it The Journal of Chemical Physics} 63, no. 6: 2479-2483. DOI:10.1063/1.431689.

\bibitem{Weinhold2} Weinhold,F.1975 "Metric geometry of equilibrium thermodynamics. II. Scaling, homogeneity, and generalized Gibbs–Duhem relations." {\it The Journal of Chemical Physics}63, no. 6 : 2484-2487., doi/10.1063/1.431635.

\bibitem{Wang} Wang, X.N. and Gyulassy, M.1991, "HIJING: A Monte Carlo model for multiple jet production in pp, pA, and AA collisions". 
{\it Physical Review D}, 44(11), p.3501.

\bibitem{P238}  Zha, et al.2018, "Coherent lepton pair production in hadronic heavy ion collisions" {\it Physics Letters B} 781 (2018), 182-186.


\end{thebibliography}
\end{document}